\documentclass[preprint,authoryear,10pt]{elsarticle}

 \usepackage{graphics}
 \usepackage{hyperref}
 \usepackage{epstopdf}
 \usepackage{float}
\usepackage[table]{xcolor}
\usepackage{amssymb}
\usepackage{amsmath}
\usepackage{placeins}
\usepackage{natbib}
\usepackage[nodots]{numcompress}
\numberwithin{equation}{section}
\numberwithin{table}{section}
\numberwithin{figure}{section}
\nocite{*}
 \biboptions{longnamesfirst,comma}

\begin{document}

\begin{frontmatter}



\title{Novel solutions toward high accuracy automatic brain tissue classification in young children}


\author{Nataliya Portman\corref{cor1}}
\ead{nataliyaportman@gmail.com}

\author{Paule-J Toussaint}
\ead{paule.toussaint@mcgill.ca} 

\author{Alan C. Evans}
\ead{alan.evans@mcgill.ca}
\ead[url]{http://www.bic.mni.mcgill.ca/~alan}

\cortext[cor1]{Corresponding author}
\address{McConnell Brain Imaging Centre, Montreal Neurological Institute, McGill University, Montreal, QC, Canada}

\begin{abstract}
Accurate automatic classification of major tissue classes and the cerebro-spinal fluid in pediatric MR scans of early childhood brains remains a challenge. A poor and highly variable grey matter and white matter contrast on T1-weighted MR scans of developing brains complicates the automatic categorization of voxels into major tissue classes using state-of-the-art classification methods (Partial Volume Estimation). Varying intensities across brain tissues and possible tissue artifacts further contribute to misclassification.
\newline
In order to improve the accuracy of automatic detection of major tissue types and the cerebro-spinal fluid in infant brains within the age range from 10 days to 4.5 years, we propose a new classification method based on Kernel Fisher Discriminant analysis (KFDA) for pattern recognition, combined with an objective structural similarity index (SSIM) for perceptual image quality assessment. The proposed method performs an optimal partitioning of the image domain into subdomains having different average intensity values and relatively homogeneous tissue intensity. In the KFDA-based framework, a complex non-linear structure of grey matter, white matter and cerebro-spinal fluid intensity clusters in a 3D (T1w, T2w, PDw)-space is exploited to find an accurate classification. Based on Computer Vision hypothesis that the Human Visual System is an optimal structural information extractor, the SSIM finds a new role in the evaluation of the quality of classification.
\newline
A comparison with state-of-the-art Partial Volume Estimation method using SSIM index demonstrates a superior performance of the local KFDA-based algorithm in low contrast subdomains and a more accurate detection of grey matter, white matter, and cerebro-spinal fluid patterns in the brain volume.
\end{abstract}

\begin{keyword}
Kernel Fisher Discriminant Analysis \sep Structural Similarity \sep brain tissue classification \sep early brain development \sep NIH Objective-2 \sep intensity variability \sep low contrast 
%% or \MSC[208] code \sep code (2000 is the default)

\end{keyword}

\end{frontmatter}

% \linenumbers

\section{Introduction}
\label{intro}
\subsection{Dataset and its segmentation challenges}
\begin{figure}
\includegraphics[width=4.8in,height=1.5in]{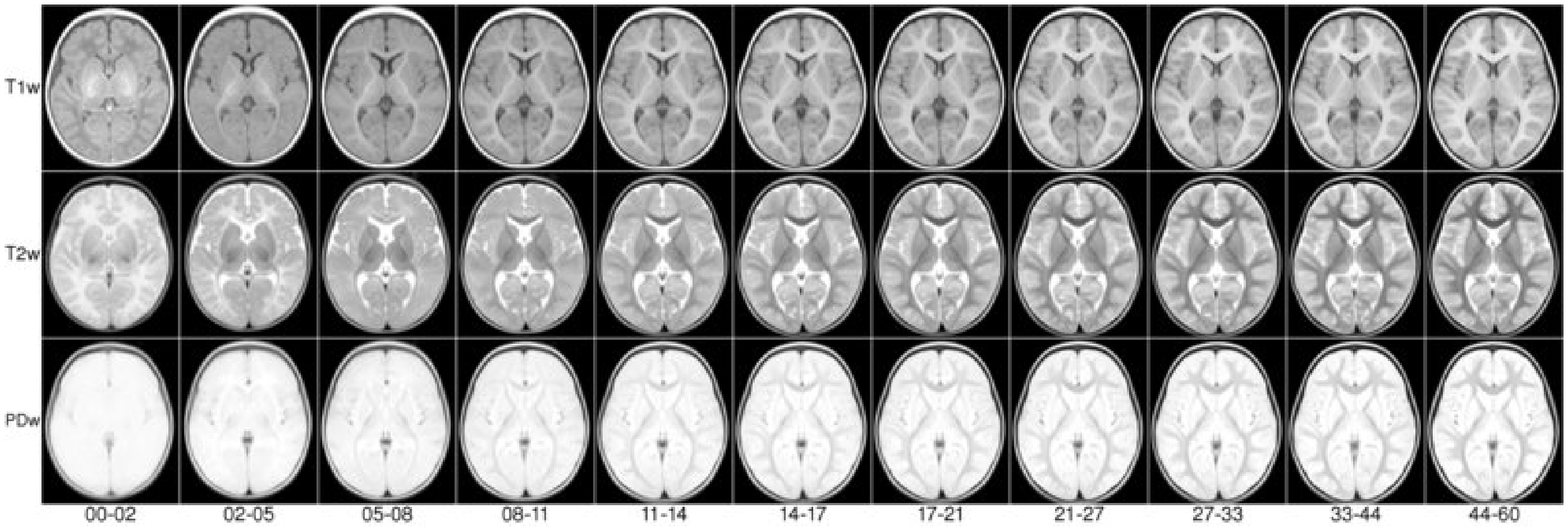}
\centering{\tiny{(Courtesy of V. Fonov. Available online at at http://www.bic.mni.mcgill.ca/ServicesAtlases/NIHPD-obj2)}}
\caption{T1-(top), T2-(centre) and PD-weighted (bottom) average atlases for important developmental age ranges (in months) for the Objective-2 MRI data.}
\label{fig1}
\end{figure}

\begin{figure}[h!]
  \centering
 \includegraphics[width=4in,height=6in]{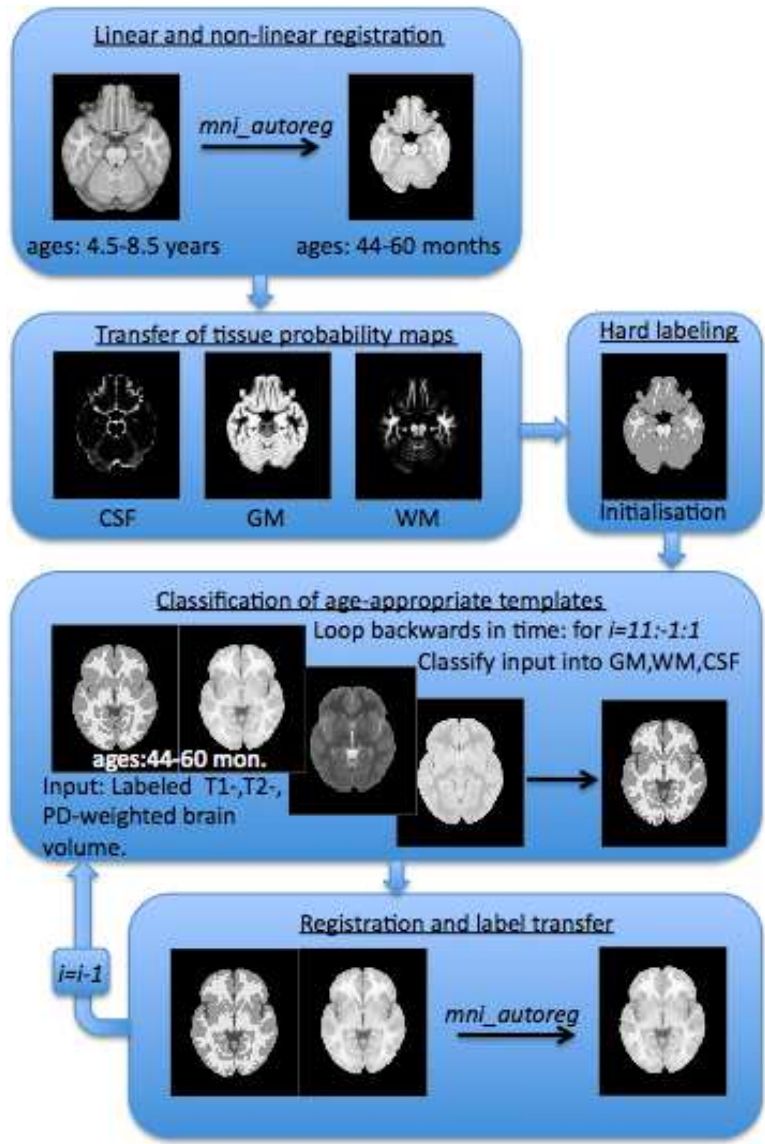}
 \caption{Flow diagram of the overall system for brain tissue classification of age-dependent pediatric Objective-2 templates.}
 \label{fig2}
 \end{figure}

This paper addresses a need for the development of automatic and accurate brain tissue classification techniques and validation of classification in MR brain images obtained during early childhood development. Namely, we consider the National Institutes of Health (NIH) pediatric database that represents the largest demographically diverse U.S. pediatric cohort with respect to ethnicity, gender, race and income level \citep{Alm2007, Evans2006}. Within the larger cohort from birth to adolescence, the "Objective-2" (O2) MR subset consists of 69 typically developing subjects in the age range from 10 days to 4.5 years scanned longitudinally. 
\newline
\cite{Alm2007} and \cite{Fon2011} have reported that the dynamic changes in intensity and brain shape with age in the O2 data pose a challenge to using general anatomical image processing pipelines such as FSL \citep{SJen2004}, CIVET \citep{MacDonald2000} and statistical analysis tools provided by SPM \citep{AshF1997}. 
\newline
There is a high variability of MR signal across times and within each tissue class brought about by rapid changes in signal intensity of grey matter and white matter (Figure \ref{fig1}). In particular, in the course of time white matter reverses in intensity  as axons become myelinated  from occipital to frontal lobes, and later from central to subcortical white matter. 
\newline
Secondly, the contrast-to-noise ratio between grey matter and white matter in the pediatric brain MRI can be as low as half of that in adult MRI. Automated state-of-the-art classification algorithms fail to delineate accurately tissue class clusters in brain subdomains with significantly overlapping intensity histograms. 
\newline 
Thirdly, due to the risk of motion in non-sedated infants during brain scanning, a trade-off between spatial accuracy and motionless acquisition time was achieved at a resolution of $1\times1\times3\: mm^3$. The MRI resolution is critical to the quality of segmentation. Thicker axial slices accentuate the partial volume artifact, a voxel that contains signal from two or three different tissues.  
\newline
There is a need for new quantitative analysis techniques for the O2 data that will accommodate their unique characteristics such as high tissue intensity variation and changing grey/white matter contrast. As these changes occur rapidly over time during brain development, analytical methods for early pediatric MRI must employ age-specific anatomical templates (Figure \ref{fig1}).
In this paper, we implement a novel KFDA-based classification method \citep{Portman2014} to automatically generate GM, WM and CSF labels (Figure \ref{fig2}) in age-dependent pediatric O2 templates. We started with the youngest brain template for ages 4.5 to 8.5 years from NIH Objective-1 (O1) pediatric database \citep{Fon2011} that represents the average anatomy closest in age to the oldest brain template for ages 44 to 60 months from the O2 database. For the initial estimation of GM, WM and CSF spatial patterns, brain tissue probability maps were transferred from the youngest O1 T1w template to the oldest O2 T1w template using \textit{mni$\_$autoreg} \citep{ColN1994}. Then, hard GM, WM and CSF labels were obtained from the transferred tissue probability maps by choosing the tissue class with a highest probability at a given brain voxel.
 \newline
Having combined the best guess on GM, WM and CSF location with multi-channel,  T1-, T2- and PD-weighted template volumes, tissue labels for the oldest O2 template were generated using the KFDA algorithm. The process can then be repeated to map these labels to the second oldest template for the age range 33-44 months and so on in a pairwise comparison to the lowest age group. 
\newline
Figure \ref{fig1} shows drastic intensity changes in WM as it reverses in intensity due to the myelination process and complicates the pairwise template diffeomorphic registration. The subcortical WM of the frontal and temporal lobes is not myelinated for the age range 5-8 months and the T1w signal of WM in these regions is isointense with GM. For ages 2 to 5 months the myelinated area spreads from the internal capsule to the occipital lobe along cortico-spinal tracts, and for an earlier age it occupies the internal capsule only. For these early childhood templates, an extension of the proposed algorithm to identification of myelinated and unmyelinated WM in addition to GM and CSF is needed, and is beyond the scope of this paper.
\subsection{Classification techniques in child brain MRI}
The development of segmentation techniques in newborn and infant brain MRI over the past decade has shown that atlas-based segmentation methodology for adult brain MR images can be applicable to child brain MRI when tuned to the specific properties of the dataset under study. There are three major child MRI segmentation frameworks, Expectation-Maximization, Registration-based method and Adaptive Label Fusion:
\newline
\textbf{Expectation-Maximization (EM) framework.}
This widely used framework assumes a global Gaussian mixture model of the distribution of major tissue class intensities \citep{VanLeem1999}. The EM method relies on spatial priors from probabilistic atlases that describe anatomical variability of the brain population and constrains classification during parameter estimation step. 
\newline
Since application of adult prior probability maps to pediatric brain atlas construction generates errors due to anatomical differences, studies have focused on generation of a probabilistic atlas that would appropriately capture brain anatomy in early childhood. In particular, age-specific atlases of GM, WM and CSF tissue classes for O1 database of children aged 4.5 to 18.5 years have been created \citep{Fon2011}.  
\newline
Various techniques for creation of population-specific brain atlases for MRI study of early brain development have been proposed \citep{Evans2012, Murg2007, Murgasova2011, Pras2004, Xue2007}. Most of these atlases are derived from small cohorts that do not capture the full range of normal anatomical variability. Furthermore, they do not cover the entire O2 developmental epoch from birth to 4.5 years of age. The creation of such standard atlas still remains a significant challenge since it requires semi-automated segmentation of large datasets.
\newline
The successful EM classification of child brain MRI is achieved at a huge computational cost associated with building an appropriate brain atlas and overcoming restrictions of the global Gaussian mixture modelling \citep{Pras2004, VanLeem1999, Wells1996, Xue2007, Toh2004, Choi1991}. In reality, voxel intensity distribution within infant brain MRI differs from a Gaussian due to partial volume effects and biological intensity variation.
\newline
 \textbf{Registration-based framework.}
Registration-based brain segmentation methods are seemingly attractive in applications to the population of young children since tissue intensity variations do not affect the segmentation results and there is no need for tissue intensity probabilistic models. A powerful technique known as ANIMAl+INSECT (developed by \citep{ColZij1999}) non-rigidly registers an adult brain atlas to an individual subject via ANIMAL while simultaneously correcting for intensity inhomogeneity in the subject. The ANIMAL procedure iteratively estimates a 3D non-linear deformation field iteratively in a multiscale hierarchy by optimal matching of both Gaussian blurred volumes and subsequent refining of the resulting displacement field. The estimated transformation is then used for tissue label transfer from the atlas to the subject. 
\newline
\citep{Murg2007} examined the suitability of a registration-based technique for segmentation of child brain MRI at ages 1 and 2 years. The method outperformed EM \citep{VanLeem1999} in segmentation of subcortical brain structures (such as the thalamus) that exhibit substantial intensity variation. However, EM was more successful in cortical areas due to the difficulty of registering complex cortical folds in atlas and subject brains.
\newline   
As with EM, a barrier to implementation of registration methods is the dependency on a pediatric brain atlas with accurate measures of spatial boundary uncertainty for tissue classes that the O2 dataset does not possess.  
\newline   
\textbf{Adaptive label fusion.}
Adaptive label fusion is a machine learning approach that estimates subject-specific tissue intensity distributions non-parametrically and therefore, captures complex distributions of voxel intensities in young child MRI \citep{Wei2009}. Given a library of newborn template MRI each containing manually selected tissue class prototypes, the tissue prototype labels of each template are transferred to the subject through the registration process. Then the conditional (on the subject intensities) probability density of the tissue labels is non-parametrically estimated based on each template prototype labels leading to candidate segmentations of the subject. The candidate segmentations are then fused using the maximum-likelihood method that computes the label weights based on how reliable the estimated candidate segmentation is.
\newline
The label fusion technique uses global intensity statistics that requires for each tissue type to produce similar intensities in different parts of the image. To manage the tissue intensity variability and partial volume effects a probabilistic atlas constructed from 15 newborn brain MRI is incorporated into the fused segmentation model \citep{War2000}. 
\newline
The label fusion approach is flexible as it may or may not be atlas-dependent. However, it requires a library of infant brain MRI templates of the appropriate developmental age with manually chosen tissue prototypes that is not yet available for the pediatric O2 dataset. 
\subsection{KFDA-based framework}
Portman \citep{Portman2014} introduced a local atlas-independent framework using modern trends in Computer Vision such as Kernel Fisher Discriminant Analysis (KFDA) for pattern recognition and structural similarity index (SSIM; \citep{Wang2004}) for perceptual image quality evaluation. In this framework, within-class tissue intensity variations are handled by optimal partitioning of the brain into overlapping subdomains having different average intensities. 
\newline
KFDA-based methods explore complex non-linear structures of the tissue clusters in the input or intensity space to find optimal separating surfaces between the class clusters. KFDA attempts to make the input data more separable by non-linearly mapping them from the input intensity space to an abstract, high-dimensional feature space where the classification is performed.
\newline 
Furthermore, the proposed method automatically identifies voxel categories such as PVE voxels and tissue class prototypes, and predicts the most likely membership of PVE voxels from the set of prototypes. 
 \newline
We aim to classify age-specific pediatric O2 templates into two major tissue classes (GM, WM) and the CSF. The classified representative templates can then be used for the classification of a subject brain MRI of a developmental age similar to the age range of the template.      
\newline
An important question that arises with classification is validation. In the absence of a ``ground truth", we assess the quality of classification via SSIM that predicts image quality as perceived by the Human Visual System (HVS) \citep{WangBovik2009} which is regarded as an optimal information extractor that seeks to identify objects in the image. As such, the HVS must automatically identify structural distortions and compensate for the nonstructural distortions (e.g., image corruption by noise). The SSIM is an objective image quality metric that simulates this functionality and computes the degree of structural similarity between reference and distorted images. It has been shown that the SSIM is well-matched to the subject ratings of image databases, and therefore, it is a good approximation to the perceived image quality  \citep{Wang2003}.
\newline
Given the objectivity and effectiveness of the SSIM we apply it for comparison of Partial Volume Estimation \citep{Toh2004} and KFDA-based classification algorithms, as well as for automatic monitoring of the quality of classification. That is, we compute the structural closeness of classified 3D brain images containing GM, WM and CSF mean intensity values with their counterparts seen in an MR image type of a highest contrast and regarded as references. For the age range from 8 to 60 months, we rely on T1w as the most informative of all MR image types. However, for an earlier age group T1w template serves as a reference only for the CSF pattern (Figure \ref{fig1}) due to a poor GM/WM contrast. For GM and WM, T2w data can be used as a reference since they exhibit higher GM/WM contrast.
\subsection{Paper organization}
In the following we will first illustrate the pipeline structure of the proposed segmentation algorithm, and discuss a need for local segmentation. We will then describe the algorithm step by step, namely, optimal brain partitioning, subsequent delineation of the CSF and WM clusters, and image stitching. 
Finally, we will report on classification results for O2 brain templates for ages from 44 to 60 months and for ages 8 to 11 months, compare the performance of PVE and KFDA methods and discuss possible improvements and further applications of our method. 
\section{Method}
\label{method}
  
\subsection{Overview}
\label{overview}
The problem of classification into GM, WM and the CSF is partitioned into two binary classification problems.
\begin{figure}[ht]
\centering
\includegraphics[width=7cm, height=12.5cm]{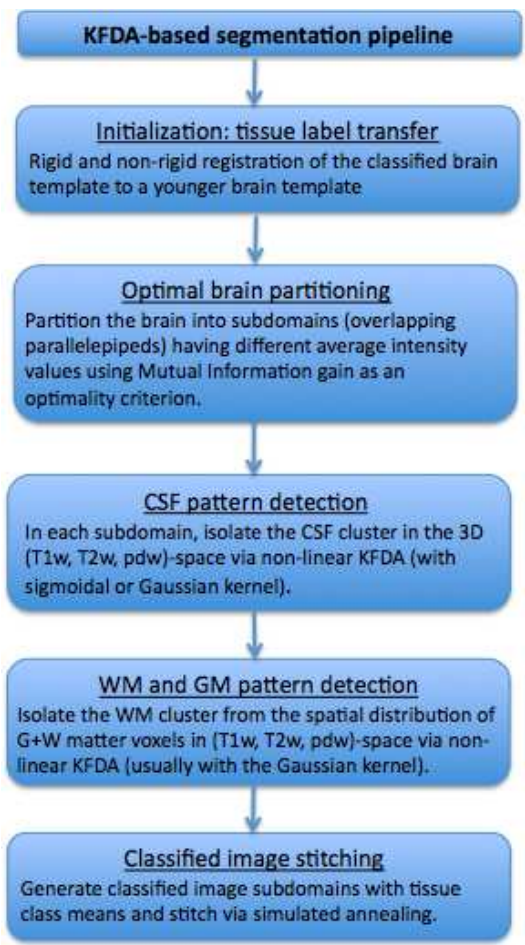}
\caption{Pipeline structure of the KFDA-based segmentation method.}
\label{fig3}
\end{figure}
First, the multichannel (T1w, T2w, pdw) image data are separated into the CSF and G+WM using a kernel function that best describes the non-linear behaviour of the CSF cluster in the 3D intensity space. We postulate that in this way we can delineate the CSF with a higher accuracy. As reported in \cite{Bouix} the major drawback of the existing automatic classification algorithms is incorrect classification of the CSF. 
\newline
Second, the G+WM data is separated into GM and WM. These are major steps of the proposed KFDA-based method whose pipeline structure is illustrated in Figure \ref{fig3}.  
\subsection{Data preprocessing}
\label{materials}
Healthy children aged 10 days to 4.5 years (n=106) were enrolled for the NIH-funded O2 MRI study of early normal brain development. In this project, T1w,T2w and pdw data were acquired approximately at quarterly time intervals \citep{Alm2007} at a 1.5 T Siemens Sonata scanner with a $1\times1\times 3$ mm spatial resolution. The data were resampled to $1$ mm$^3$ grid using tri-cubic interpolation. The brain image quality control was applied to the MR data that reduced the original sample to 69 subjects. T1w, T2w and PDw average atlases have been created for these age ranges \citep{Fon2011}. All images were N3 non-uniformity corrected \citep{SledZij1998}, linearly normalized to have the same intensity range as the ICBM152 template by a linear histogram scaling \citep{Nyul1999} and registered to the ICBM152 stereotaxic space using MINC \textit{mni\_autoreg} non-rigid registration tool \citep{ColN1994}. Brain masks were created using BET from the FSL package \citep{Smith2002}. Brain masking was applied to all templates to remove non-brain tissue prior to classification.  
\subsection{Why brain partitioning?}
\label{why}
\begin{figure}[h!]
\centering
\includegraphics[width=5.5cm, height=3.5cm]{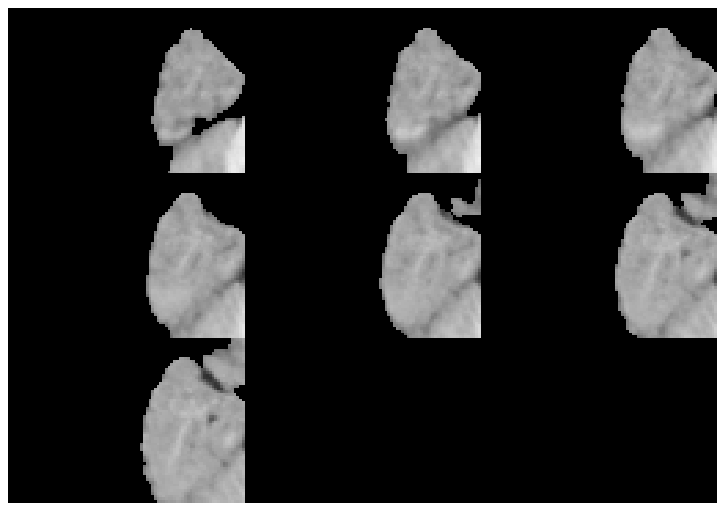}
\includegraphics[width=5.5cm, height=3.5cm]{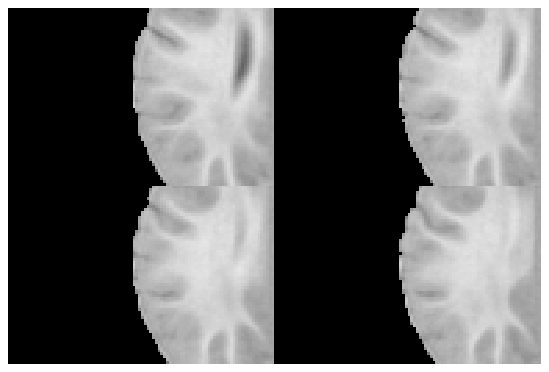}
\\
(a) \hspace{4.5cm} (b) 

\includegraphics[width=6cm, height=4.4cm]{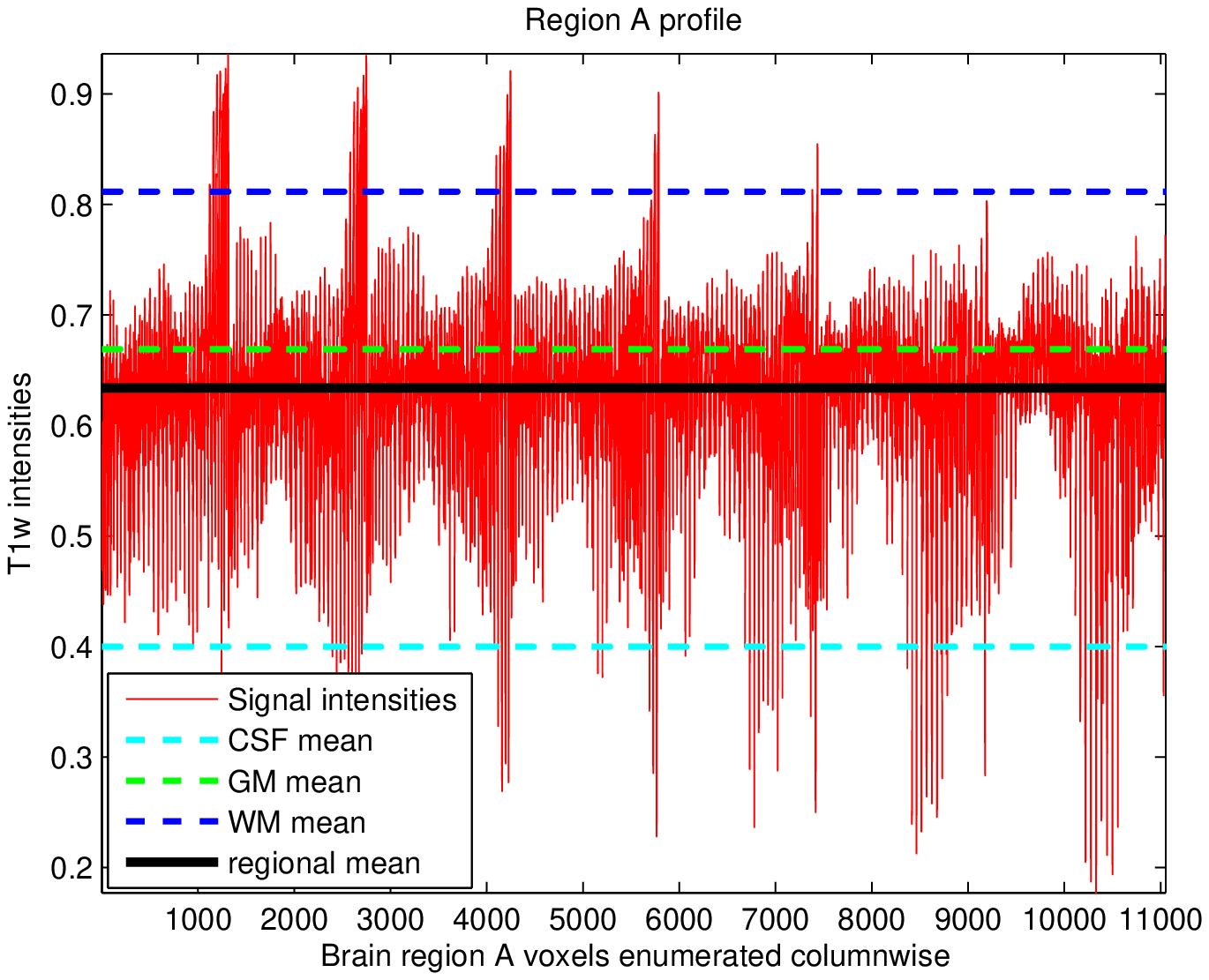}
\includegraphics[width=6cm, height=4.4cm]{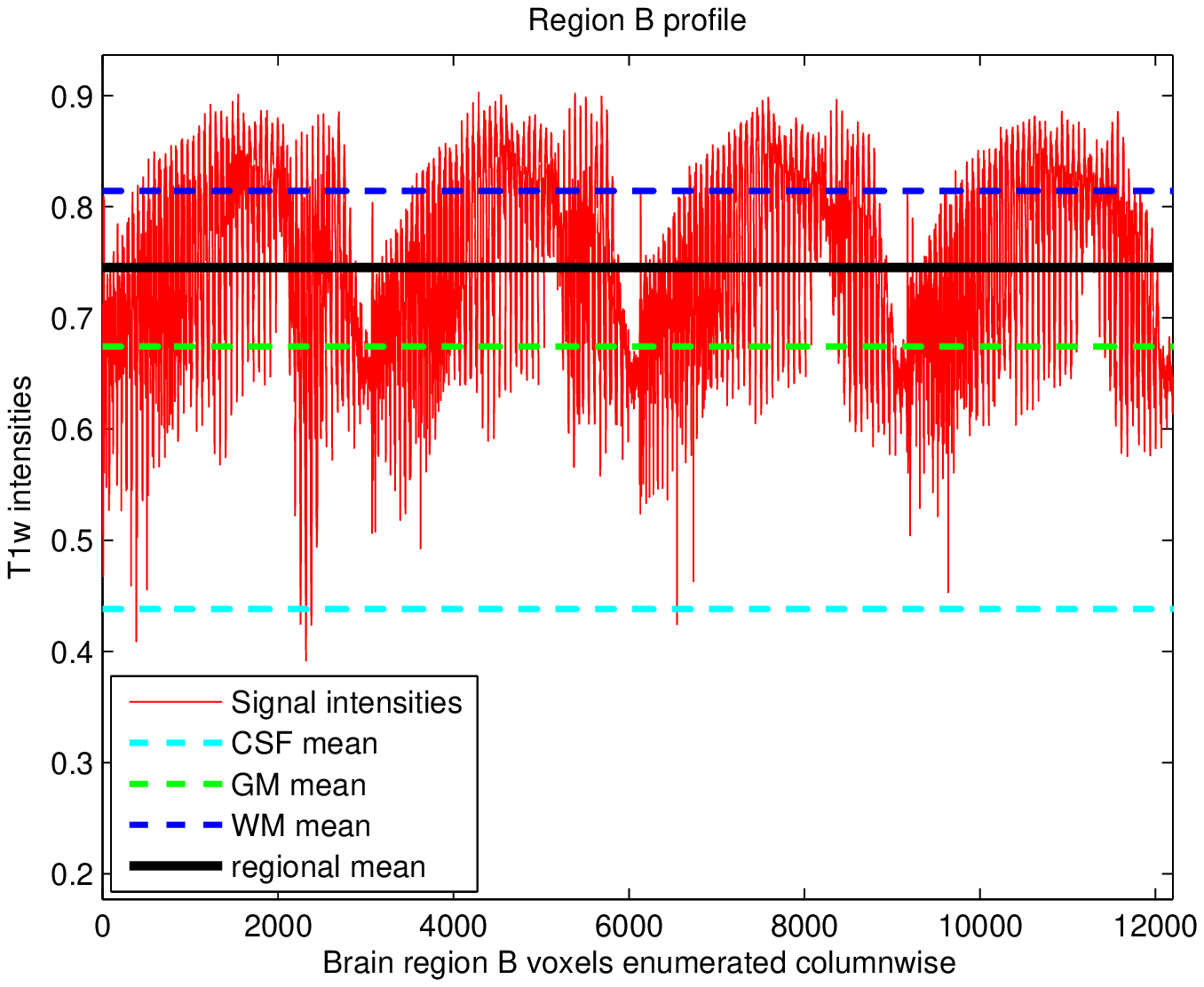}
\\
(c) \hspace{4.5cm} (d) 
\caption{(a),(b) T1-weighted template subdomains A and B from the T1w template for ages 8 months to 11 months, (c),(d) Image profiles (columnwise recordings of T1w intensity voxels of the interior brain) of subdomains A and B.}
\label{fig4}
\end{figure}

\begin{figure}[h!]
\centering
\includegraphics[width=5.5cm, height=3.5cm]{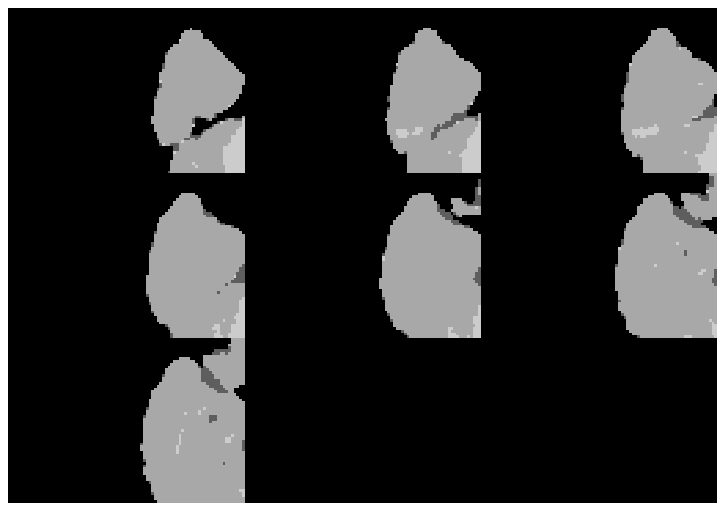}
\includegraphics[width=5.5cm, height=3.5cm]{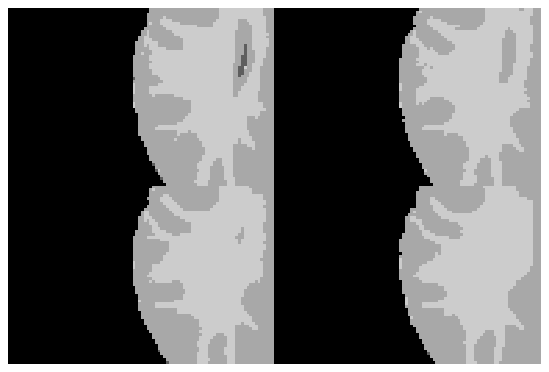}
\\
\hspace{0.6cm}(a) \hspace{4cm} (b) 
\caption{(a),(b) Global PVE-based classification in subvolumes A and B given in Figures \ref{fig4}.a and \ref{fig4}.b.}
\label{fig5}
\end{figure}

We examined infant MR image data and made a simple but important observation that led to the optimal brain partitioning algorithm presented below. 
For the O2 T1-weighted template for ages 8 months to 11 months with GM, WM and CSF probability maps obtained by global PVE we performed hard labelling into GM, WM and CSF by choosing the highest probability class at each brain voxel. We took samples of about 10,000 voxels contained in disjoint parallelepiped-like brain subdomains and plotted the corresponding T1w intensity profiles along with the tissue and subdomain intensity means as seen in the low panel of Figure \ref{fig4}. 
 \newline
 For simplicity, we illustrate a variable behaviour of these brain data samples using examples of two brain subvolumes denoted by A and B, respectively, and shown in the upper panel of Figure \ref{fig4}. Notice by comparison of diagrams \ref{fig4}.c and \ref{fig4}.d that the tissue means are approximately the same in their values as a result of global estimation of the probability distribution of tissue classes across the entire brain. However, subdomains A and B differ in their overall mean intensities.
The observations of various brain data samples suggest that there is more uncertainty in identification of tissue boundaries in subvolumes with lower overall means. Such samples usually contain lower intensities of both GM and WM classes. 
\newline
Figure \ref{fig5}.a shows the classified image subdomain A with underestimated WM and the CSF. In this subdomain, WM has a range of intensities appearing bright in the central part of the brain and fading along WM fibers. A low intensity signal of myelinated WM can be seen in the temporal lobe. As expected, it is not detected by the global PVE due to low GM/WM contrast in the temporal brain region. Similarly, a low GM/CSF contrast causes misclassification of the CSF into GM. We conclude that a lower subdomain intensity mean is an indicator of a ``problematic'' subvolume with a lower contrast. Conversely, the global classification yields quite accurate WM, GM and CSF patterns in subdomain B with a higher mean intensity as observed by comparison to GM, WM and CSF that we can visually identify in T1w data of subdomain B given in Figure \ref{fig4}.b (Figure \ref{fig5}.b).
We then applied a KFDA-based algorithm \citep{Portman2013} locally to classify low intensity image subdomain A into GM, WM and the CSF. We used the PVE-based classification given in Figure \ref{fig5}.a to initialize our algorithm. 
\begin{figure}[h!]
\centering
\includegraphics[width=5.5cm, height=3.5cm]{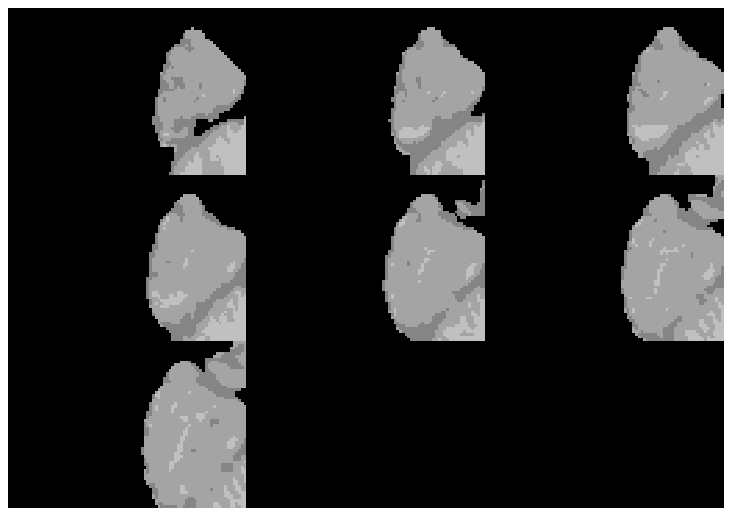}
\includegraphics[width=5.5cm, height=3.5cm]{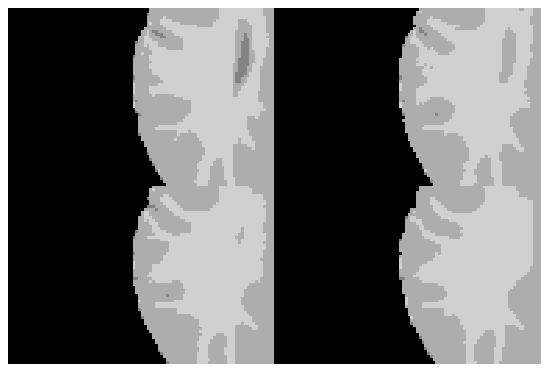}
\\
\hspace{0.6cm}(a) \hspace{4cm} (b) 
\caption{(a),(b) Local KFDA classification of subvolumes A and B given in Figures \ref{fig4}.a and \ref{fig4}.b.}
\label{fig6}
\end{figure} 
Figure \ref{fig6}.a shows a significant improvement in tissue identification over the global PVE. The local KFDA approach reveals more complex WM structure with WM streaks dispersing from the central part of the brain and the presence of more myelinated WM in the temporal lobe. In subvolumes with high intensity means (consisting of mostly WM) the WM pattern appears no different from the one detected by the global PVE as seen in Figure \ref{fig6}.b. 
\newline
Furthermore, we determined experimentally that compared to the global KFDA \citep{Portman2013} the local KFDA classification is capable to capture spatial tissue patterns with a higher accuracy. 
A comparison of tissue classifications obtained by global and local KFDA methods in the O2 brain template from 44 to 60 months of age is shown in Figure \ref{fig7}. Observe that  WM fibers in the posterior brain are detected by the local KFDA approach and omitted by the global one. 
\begin{figure}[h!]
\centering
\includegraphics[width=10cm, height=3.4cm]{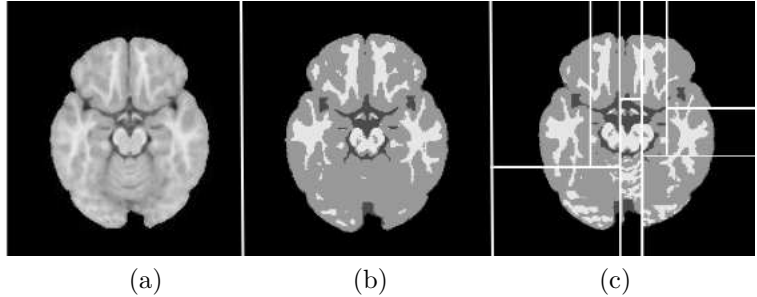}
\\
(a) \hspace{2.3cm} (b)\hspace{2.8cm}(c)
\caption{(a) T1-weighted template for the age range 44-60 months, (b) Global KFDA classification, (c) Local KFDA classification} 
\label{fig7}
\end{figure}
Therefore, we tackle intra-class MR signal variability by splitting the brain into parallelepiped-like subdomains having different average T1w intensity values. In this way, we are able to control the quality of classification locally, namely, in subdomains with low GM/WM or/and GM/CSF contrast.   
 \subsection{Optimal brain partitioning}
\label{part}
\begin{figure}[h!]
\centering
\includegraphics[width=7cm,height=3.5cm]{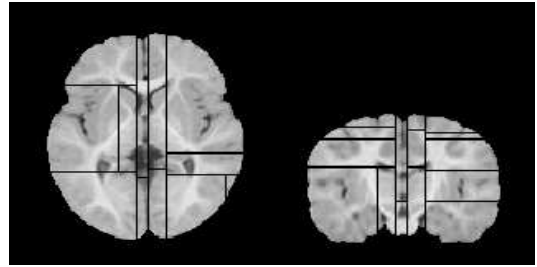}\\
(a)\\
\vspace{0.1cm}
\includegraphics[width=7cm,height=3.5cm]{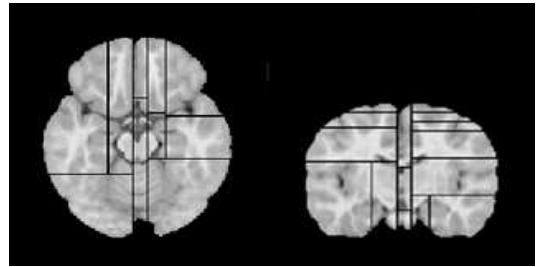}\\
(b)
\caption{(a) Transversal and coronal views of O2 T1w template brain for ages  8 to 11 months optimally partitioned into 40 subdomains, (b) Transversal and coronal views of O2 T1w template brain for ages 44 to 60 months optimally partitioned into 22 subdomains.}
\label{cnrcsf}
\end{figure}

\begin{figure}[h!]
\centering
\includegraphics[width=11cm,height=7cm]{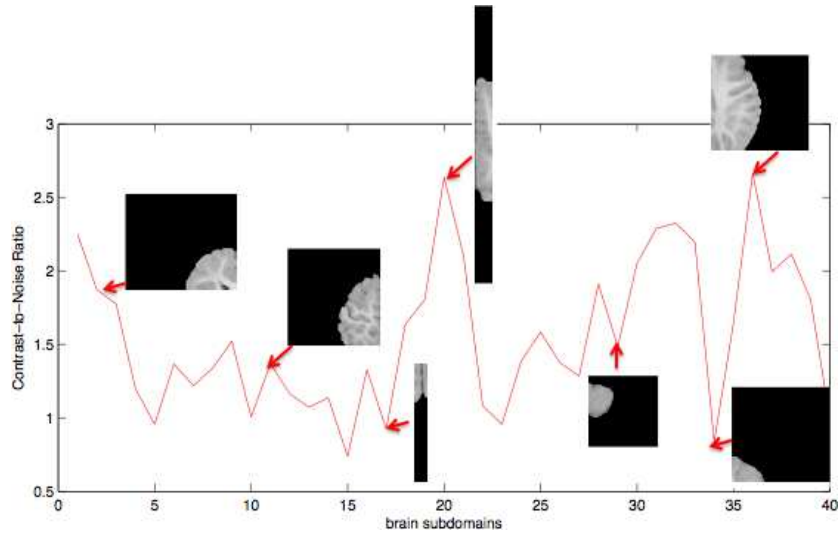}
\caption{Local dependency of GM/WM contrast-to-noise ratio across the T1w template brain for ages 8 to 11 months.}
\label{var}
\end{figure}

The algorithm for 3D brain image partitioning used here is based on maximization of mutual information (MI) between the histogram bins of the T1w image and the subdomains of the partitioned image. The algorithm progressively subdivides the 3D image into subdomains using binary space partitioning \citep{Rigau2004}, that is, each subdomain is subdivided into two subdomains with coronal, sagittal or axial planes according to the maximum MI gain. At each partitioning step, the MI optimality criterion finds a planar cut separating each parallelepiped-like volume into two subvolumes whose intensities best match the histogram bins (splitting the data into two non-overlapping intensity ranges). Thus, the obtained subvolumes will have different average intensity values.    
\newline
To maintain the continuity of the classified image subdomains across their boundaries, we added two slices to each planar boundary of each subdomain thus creating an overlap of 4 slices between adjacent brain subdomains.
\newline
Examples of the optimally partitioned brain templates for age groups 8 to 11 months and 44 to 60 months are shown in Figures \ref{cnrcsf}.a-b. Since the partitioning algorithm is intensity-based, different brain regions and their total number can be obtained for different individual brain scans. 
The optimal brain partitioning algorithm allows to identify lower intensity  regions. The computed GM/WM contrast-to-noise ratio (CNR) in image subdomains of the template for ages 8 to 11 months show that GM/WM CNR takes lower values in the posterior brain and higher values in the central part of the brain (Figure \ref{var}). 

\subsection{KFDA implementation}
\label{kfda}
The fundamental idea of KFDA  is to exploit naturally occurring class structures in the input space to find an accurate non-linear separating boundary. 
We performed KFDA in two steps.\\
\textit{Step1: Classification into the CSF and G+WM.} 
Through extensive experimentation we determined that the sigmoidal kernel function $K(\vec{I}_m, \vec{I})=\tanh(a({\vec{I}_m}^T \cdot \vec{I})+b)$ (with $a$ and $b$ being user-specified parameters) is the best choice for delineation of the CSF. 
 Since KFDA computes an optimal decision surface between the CSF and G+WM it easily identifies PV voxels that lie near or on the boundary between the classes.
\begin{figure}[h]
\centering

\includegraphics[width=5cm,height=3.2cm]{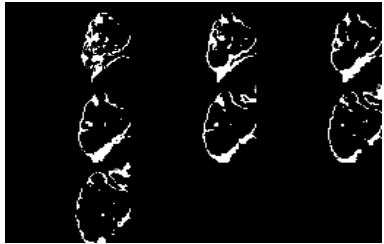}
\caption{ Overlapping set in the stereotaxic space.}
\label{csf3}
\end{figure}
A visualization of the overlapping set (coloured in black and green) in the stereotaxic space given in Figure \ref{csf3} shows that the overlapping voxels are located within the boundary regions between G+WM and the CSF. We recognize a particularly problematic brain area around the sulcus where the CSF is usually poorly detected due to the presence of PVE. Therefore, it is reasonable to assume that the overlapping voxels contain a mix of both tissue classes. 

\textit{Step2: Classification into GM and WM.}
\newline
Having delineated the CSF we classify G+WM into GM and WM. Experiments with various kernel functions showed that the Gaussian radial basis function $K(\vec{I}_i, \vec{I})=\exp\left(-\frac{(\vec{I}_i-\vec{I})^T (\vec{I}_i-\vec{I})}{2\sigma^2}\right)$ is the best choice to model the non-linear structure of WM and GM clusters. 

\subsection{Stitching of brain subdomains}
\label{stitch}
\begin{figure}[h!]
\centering
\includegraphics[width=5cm,height=4.5cm]{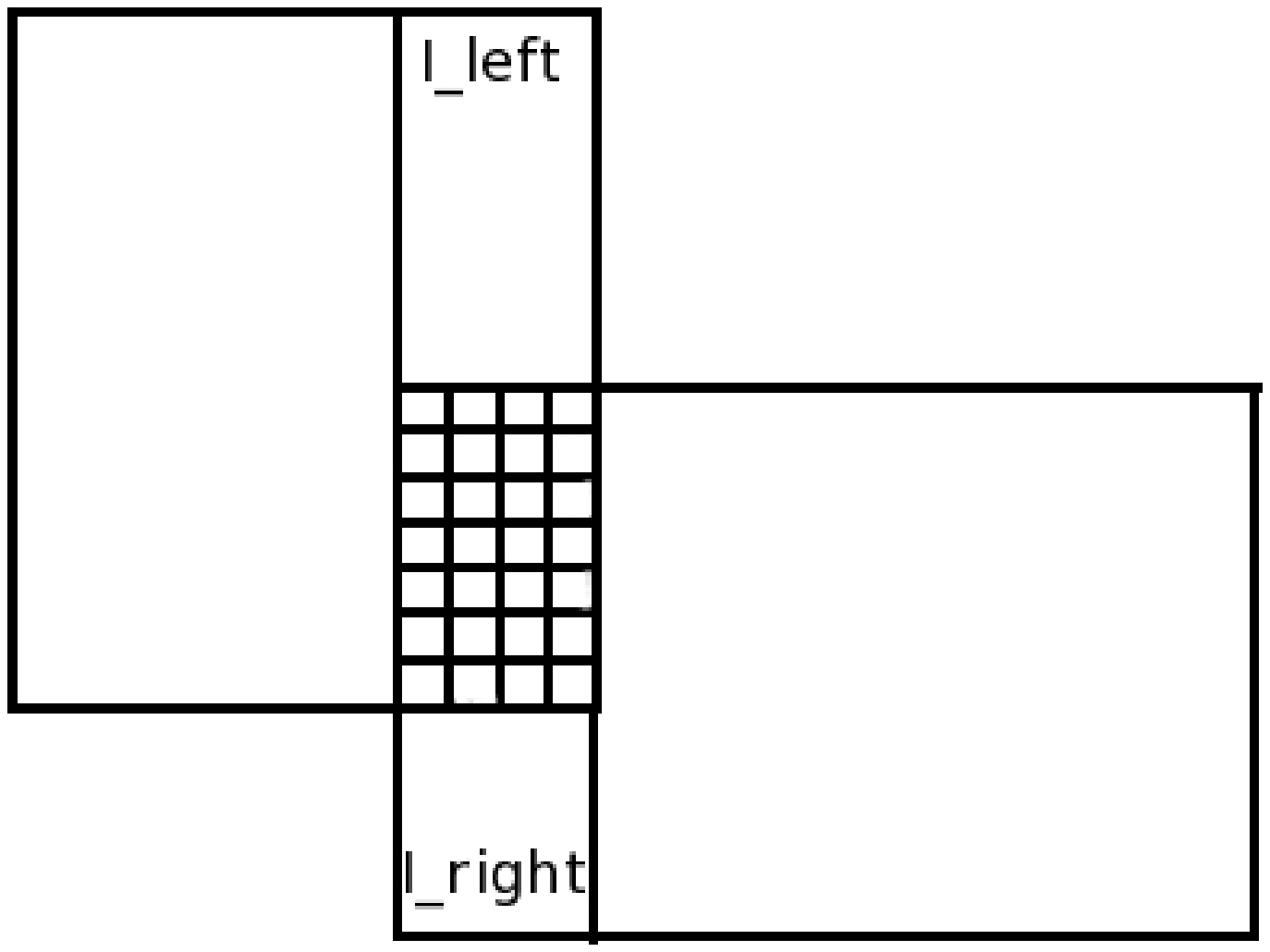}
\includegraphics[width=5cm,height=4.5cm]{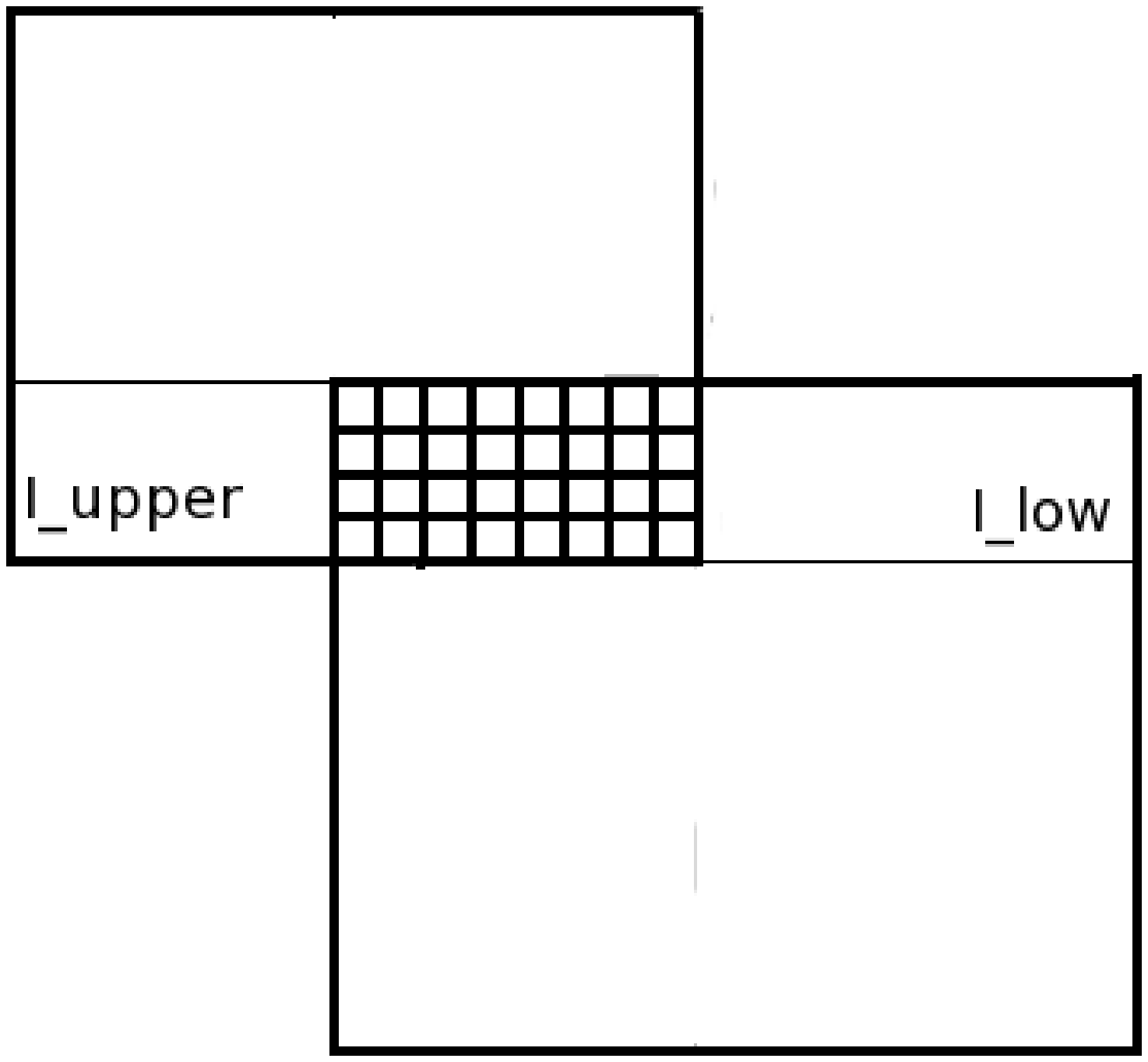}\\
(a) \hspace{3.2cm} (b) \vspace{0.3cm}\\
\includegraphics[width=9cm,height=3.8cm]{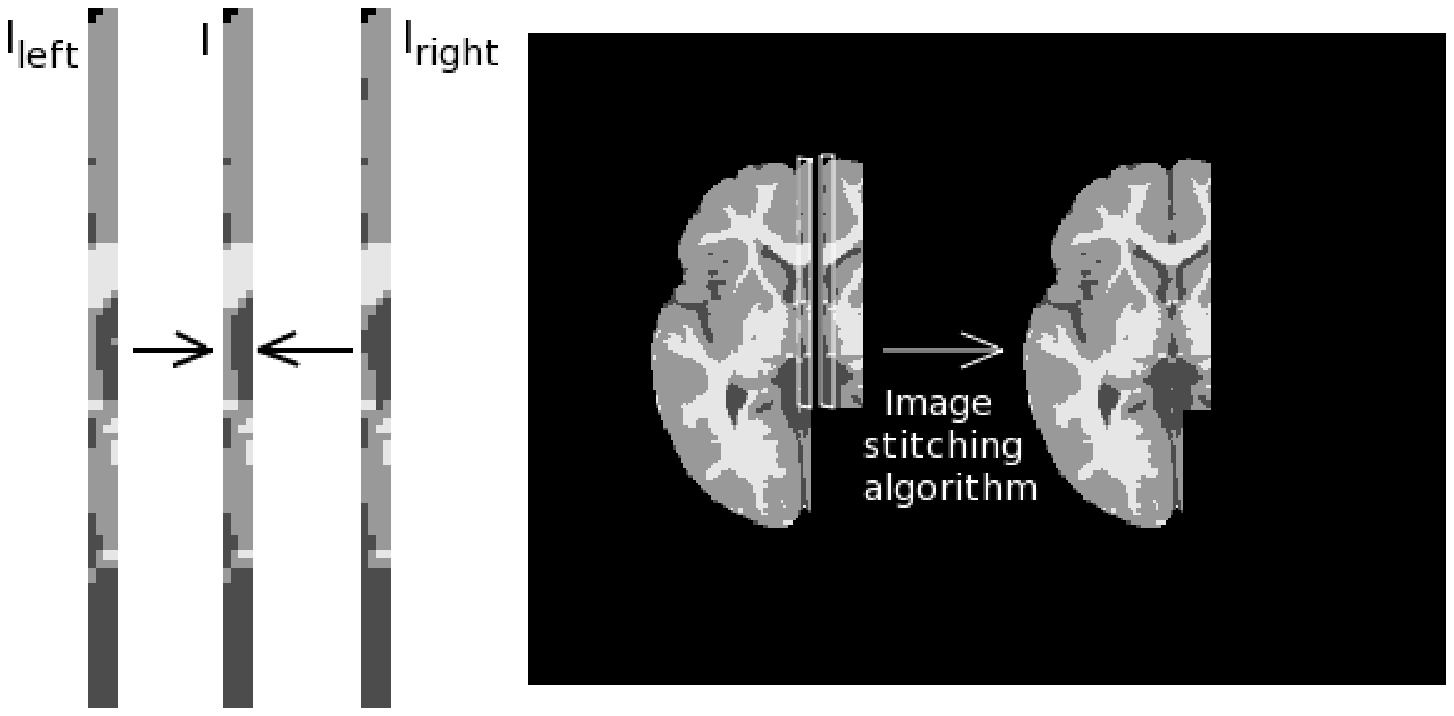}\\
(c) \vspace{0.3cm}\\
\includegraphics[width=9cm,height=6.5cm]{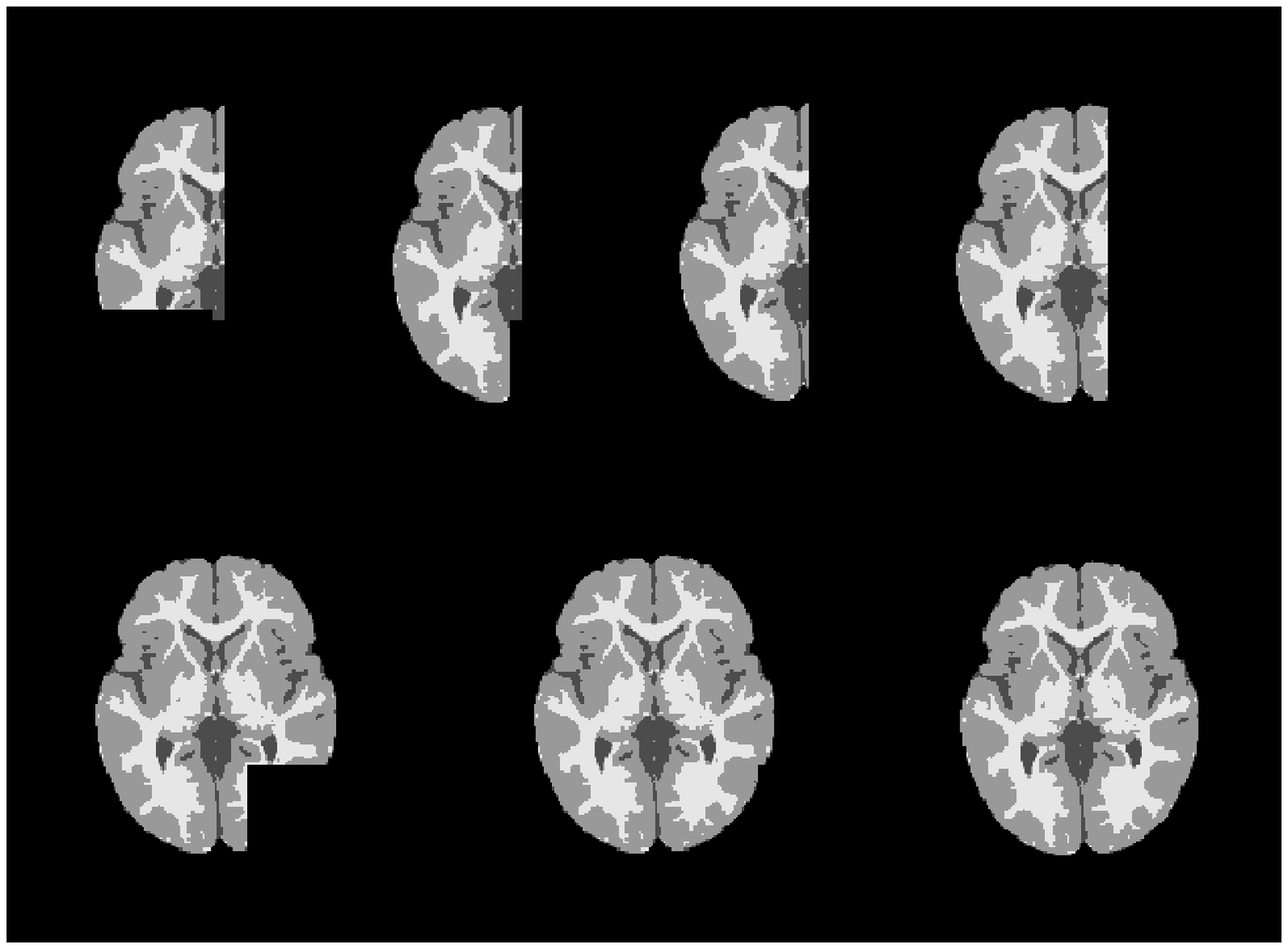}\\
(d)
\caption{(a) Left and right image region overlap, (b) upper and lower image region overlap, (c) Maximum a posteriori estimate of the label configuration $I$ in presence of observations $(I_r,I_l)$, (d) Sequental steps of the stitching algorithm in direction from top left to bottom right.}
\label{st}
\end{figure}

We applied a Simulated Annealing technique to stitch the local classifications together into a cohesive global picture of the classified brain. First of all, for each brain slice we collected the constituent subimages. Each pair of overlapping subimages contained a joint image region $I_s$ of size $4 \times ncols$ or $nrows \times 4$ that is to be optimally estimated from two sets of observations $I_l$, $I_r$ or $I_{up}$, $I_{low}$ as illustrated in Figures \ref{st}.a-b. 
We implemented the iterative Simulated Annealing (SA) algorithm \citep{Grenander1996} to find the most probable joint region for each pair of overlapping classified subimages in a brain slice. 
 \newline
An example of SA application is shown in Figure \ref{st}.c. The leftmost columns of $I_r$ and $I_l$ appear slightly different in presence of the CSF and GM.The rightmost columns of $I_l$ and $I_r$ only differ in the value of a single central pixel. The optimal labelling $I$ preserves the label configuration of its first and last columns as they appear in their respective overlapping regions $I_l$ and $I_r$. 

We started with the upper left subimage and stitched its neigbouring subimages vertically and horizontally. Then for each of the neigbouring subimages we identified overlapping ones in both horizontal and vertical directions, and glued them together using the SA algorithm. By progressive stitching of constituent image parts we assembled the entire brain slice as illustrated in Figure \ref{st}.d. The proposed SA algorithm yields seamless estimates of the joint regions.    

\section{Experiments and results}
\label{res}

\subsection{Pediatric brain template for ages 44 to 60 months}
\begin{figure}[h]
\centering
\includegraphics[width=3.5cm,height=4cm]{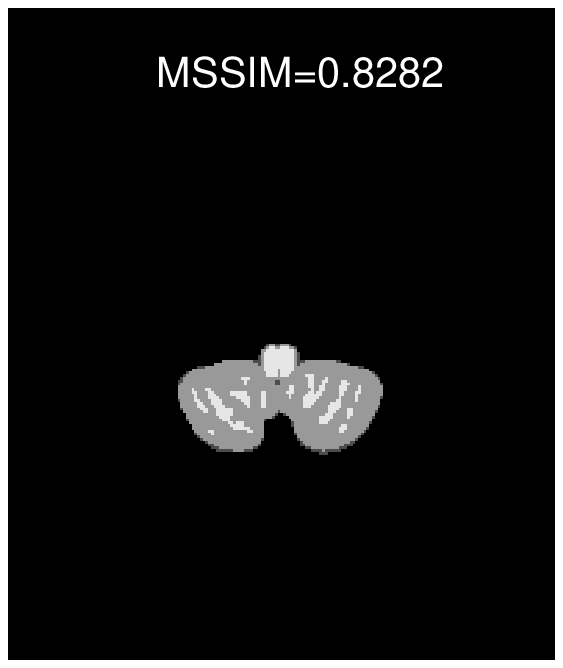}
\includegraphics[width=3.5cm,height=4cm]{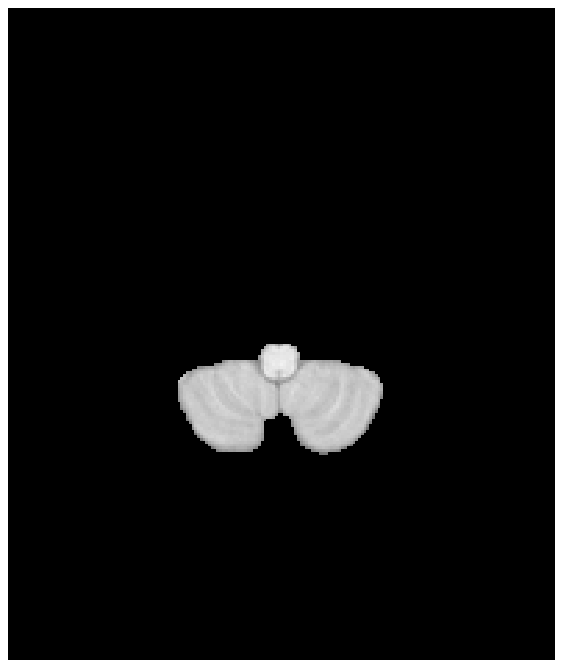}
\includegraphics[width=3.5cm,height=4cm]{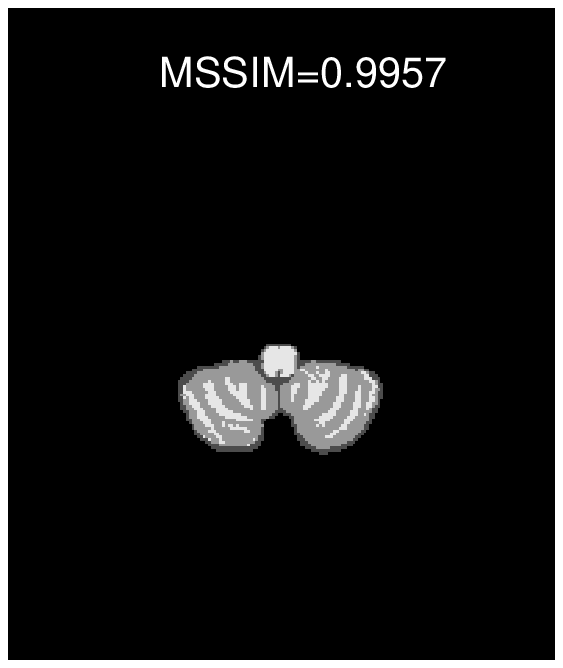}\\
(a)\hspace{3cm}(b)\hspace{3cm}(c)\\
\vspace{0.2cm}
\includegraphics[width=3.5cm,height=4cm]{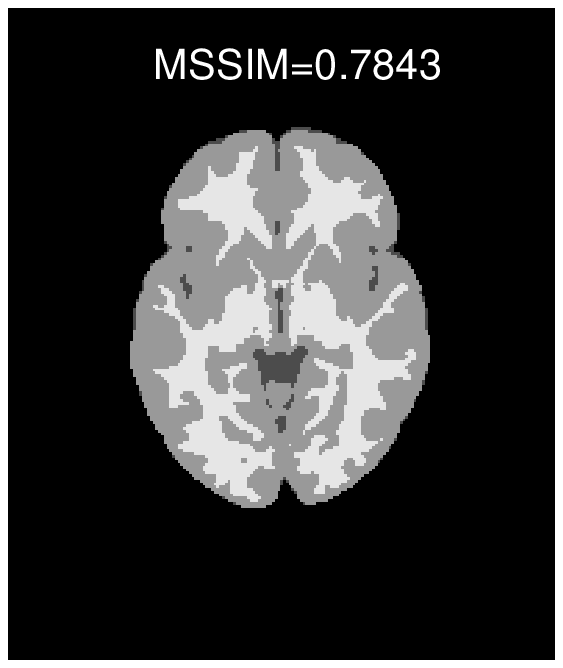}
\includegraphics[width=3.5cm,height=4cm]{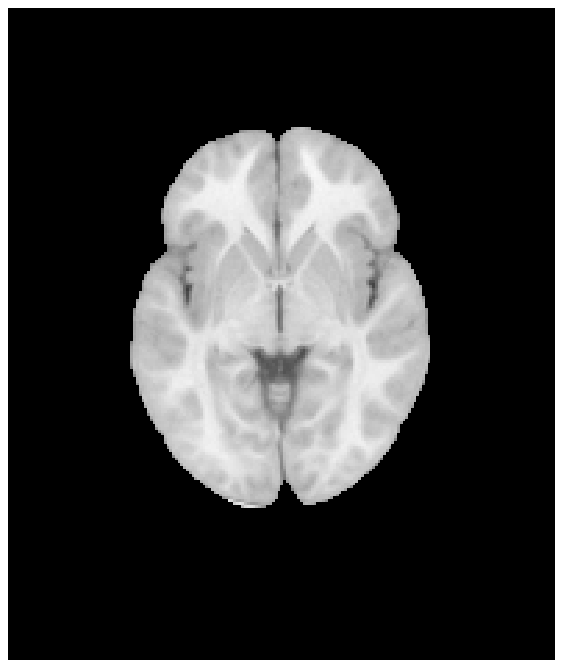}
\includegraphics[width=3.5cm,height=4cm]{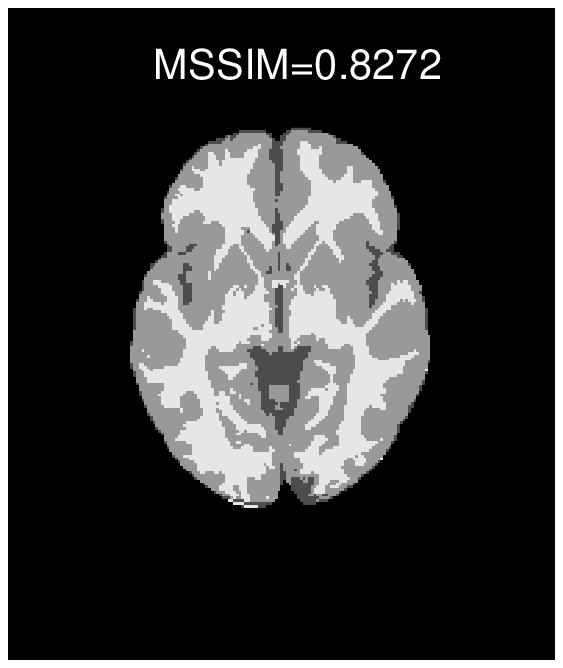}\\
(d)\hspace{3cm}(e)\hspace{3cm}(f)
\caption{Axial slices of the brain template, (a), (d) initial classification (by label transfer from an older brain), (b), (e)  T1w slices, (c), (f) KFDA classification.}
\label{sls}
\end{figure}

We applied the proposed segmentation pipeline to the O2 3D brain template for ages 44 to 60 months with initialization obtained by a label transfer from the brain template for ages 4.5 to 8.5 years.  Figure \ref{sls} illustrates the performance of KFDA using classified brain slices. 
The significant improvement in WM pattern detection was observed in the cerebellum (Figure \ref{sls}.c) and validated using MSSIM (Figure \ref{sls}.b). There is a remarkable closeness of the stripe-like WM pattern identified by KFDA to its poorly visible counterpart in a low contrast T1w brain slice.
\newline
The comparison of KFDA and initial classifications of the central axial slice demonstrates the adaptability of KFDA to individual brain tissue intensity distributions (Figures \ref{sls}.d-f). The initialization derives from the labelling of the older brain closest in age to the brain template under study. With KFDA, we obtained more accurate CSF detection, in particular in areas between the insula and precentral gyrus where the initial CSF is underestimated. The anterior horns of the lateral ventricles and the third ventricle that are not present initially in the CSF are observed in the T1w input and captured by KFDA.
\FloatBarrier
 Also, left and right frontal lobes appear to be separated by the CSF and are detected by KFDA. 
\newline
Contours of the WM in the frontal lobe are more complex, having more visible cusps in the T1w image Figure \ref{sls}.f). However, in the left frontal lobe, the WM is overestimated as it spreads into the middle frontal gyrus. WM fibers passing between the caudate nucleus and putamen appear connected in both the reference (T1w) and KFDA-classified images.
\newline
There are WM fibers separated from the main fiber bundle in the anteromedial temporal lobe with KFDA classification that are in agreement with the fading T1w signal intensity in this area of the brain. However, in the initialization generated by a label transfer from an older brain template they appear to be a part of the whole WM.
Overall, KFDA tends to accurately reproduce the segmented brain anatomy according to its natural appearance in the input MR images. 

\subsection{Pediatric brain template for ages from 8 months to 11 months}
\begin{figure}[h!]
\centering
\includegraphics[width=2.8cm,height=3.5cm]{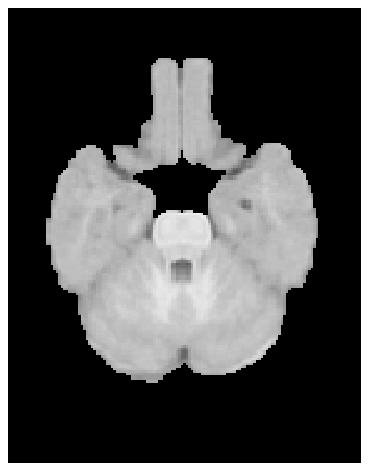}
\includegraphics[width=2.8cm,height=3.5cm]{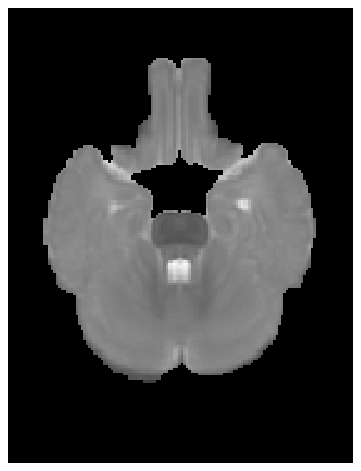}
\includegraphics[width=3.2cm,height=1.8cm]{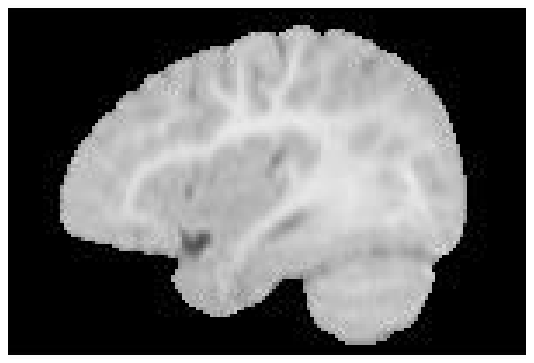}
\\
(a)\hspace{2.5cm}(b)\hspace{2.8cm}(c)\\
\vspace{0.2cm}
\includegraphics[width=2.8cm,height=3.5cm]{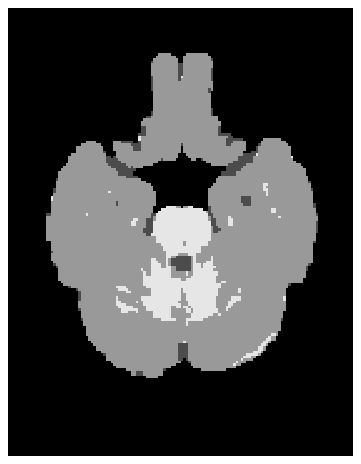}
\includegraphics[width=2.8cm,height=3.5cm]{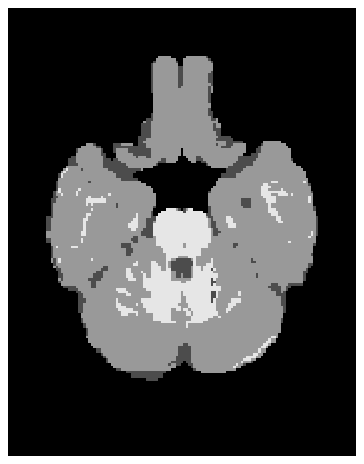}
\includegraphics[width=3.2cm,height=3.5cm]{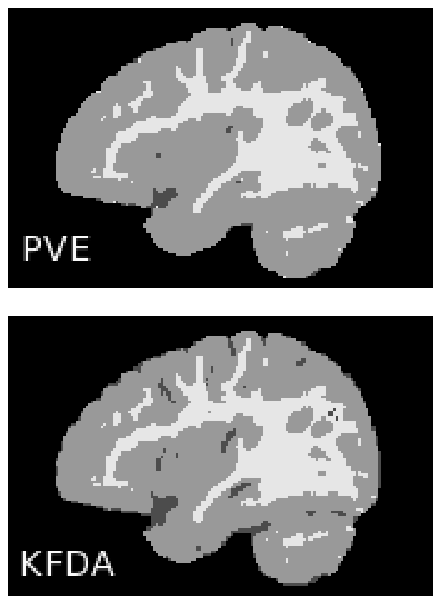}
\\
(d)\hspace{2.5cm}(e)\hspace{2.8cm}(f)\\
\caption{(a) T1w and (b) T2w axial slices from the brain template (8-11 months) and the corresponding (d) PVE and (e) KFDA classifications, (c) T1w sagittal slice and (f) its PVE and KFDA classification.}
\label{sl1}
\end{figure}

\begin{figure}[h!]
\centering
\includegraphics[width=2.8cm,height=3.5cm]{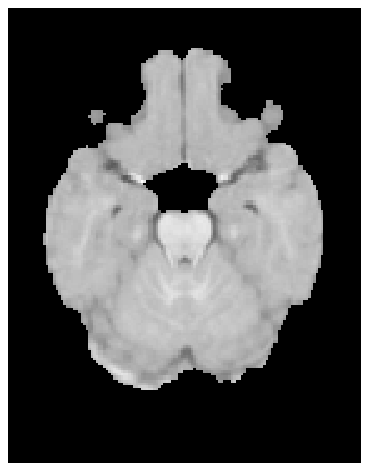}
\includegraphics[width=2.8cm,height=3.5cm]{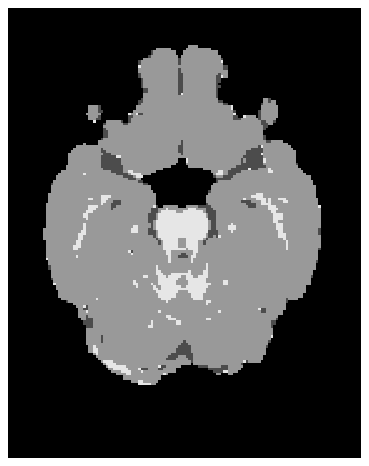}
\includegraphics[width=2.8cm,height=3.5cm]{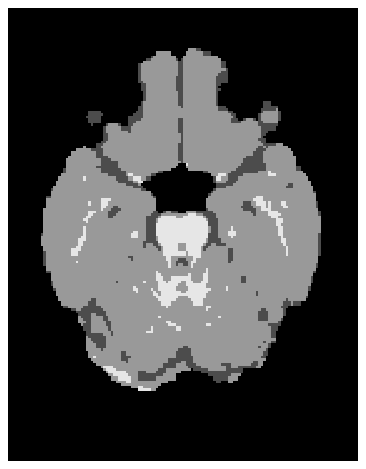}\\
(a)\hspace{2.5cm}(b)\hspace{2.5cm}(c)\\
\vspace{0.2cm}
\includegraphics[width=2.8cm,height=3.5cm]{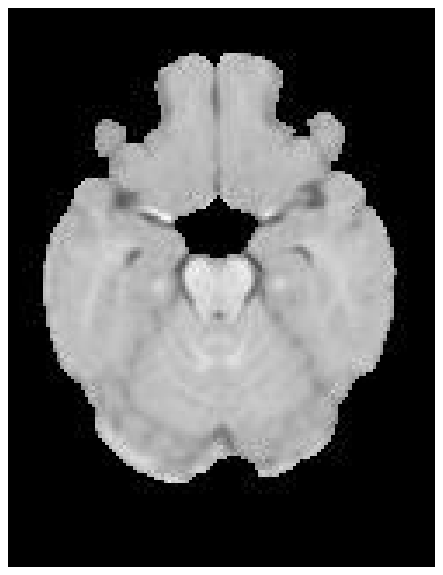}
\includegraphics[width=2.8cm,height=3.5cm]{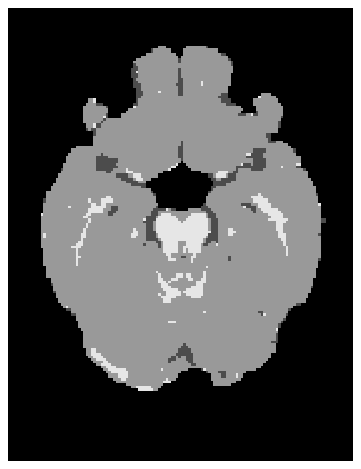}
\includegraphics[width=2.8cm,height=3.5cm]{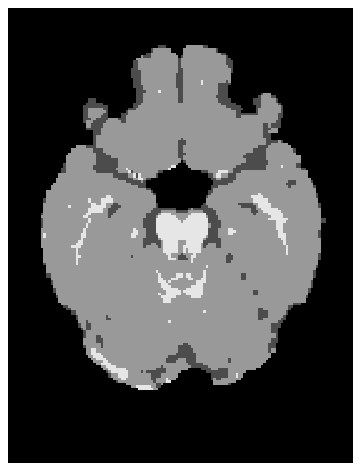}\\
(d)\hspace{2.5cm}(e)\hspace{2.5cm}(f)
\caption{(a) T1w axial slice 30 from the brain template (8-11 months) and its (b) PVE and (c) KFDA classifications,
(d) T1w axial slice 31 and its (e) PVE and (f) KFDA classifications.}
\label{sl2}
\end{figure}

We applied the local KFDA-based method to segment the O2 brain template for ages 8 to 11 months initially classified into GM, WM and the CSF using PVE. The proposed approach leads to a significant improvement in the CSF detection throughout the brain almost doubling the initial CSF volume. The initial CSF cluster determined by PVE consisted of 26717 voxels and has increased to 53681 voxels with application of KFDA. 
\newline
The comparison of KFDA and PVE classified results clearly demonstrates that CSF is underestimated in PVE classification (Figures \ref{sl1} and \ref{sl2}). KFDA reveals complex CSF patterns in the occipital lobe and on the border between cerebellum and temporal lobes as seen in Figures \ref{sl2}.c and \ref{sl2}.f. However, it does not capture the CSF completely on the border between the cerebellum and the left temporal lobe (Figures \ref{sl2}.c and \ref{sl2}.f) due to the less pronounced intensity differences between GM and the CSF in the left hemisphere.
\newline
Figure \ref{sl1}.e shows that in contrast to PVE, KFDA detects most WM seen in low-contrast temporal lobes and lower intensity WM streaks that stretch out from the central part of the cerebellar WM (Figure \ref{sl1}.a). This example demonstrates the capability of the proposed method to identify tissue patterns structurally similar to their counterparts that are visible in low contrast subdomains of T1w images.
\FloatBarrier

\section{Discussion}
\label{disc}
\begin{figure}[h!]
\centering
\includegraphics[width=2.8cm,height=3.3cm]{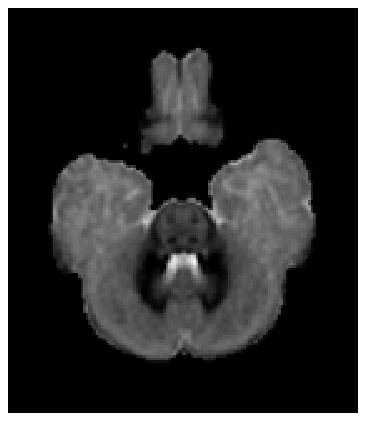}
\includegraphics[width=2.8cm,height=3.3cm]{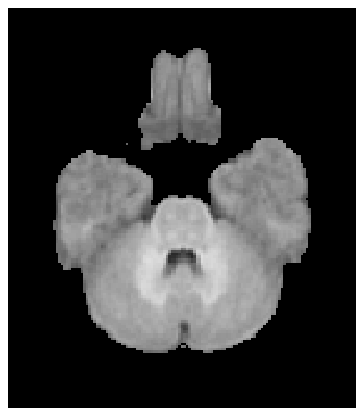}
\includegraphics[width=2.8cm,height=3.3cm]{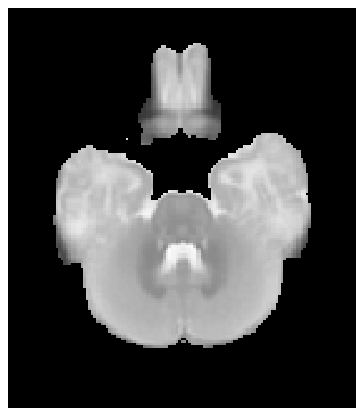}
\\
(a)\hspace{2.5cm}(b)\hspace{2.6cm}(c)\\
\vspace{0.2cm}
\includegraphics[width=2.8cm,height=3.3cm]{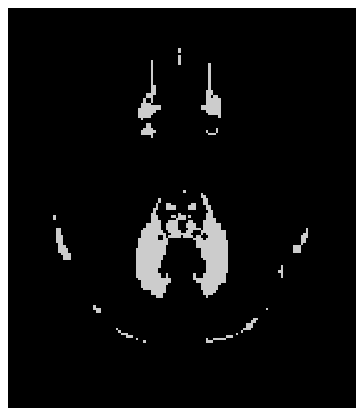}
\includegraphics[width=2.8cm,height=3.3cm]{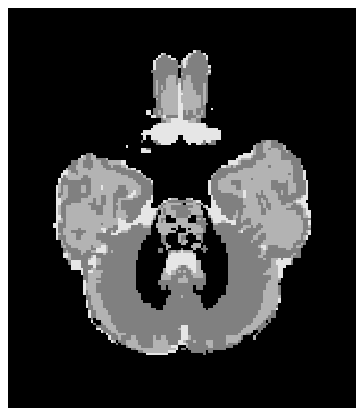}
\includegraphics[width=2.8cm,height=3.3cm]{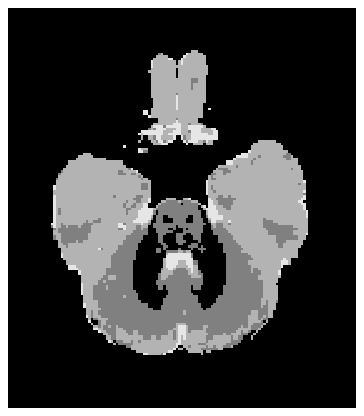}
\\
(d)\hspace{2.5cm}(e)\hspace{2.6cm}(f)\\
\caption{(a) Difference PDw$-$T1w, (b) T1w and (c) T2w axial slices from an O2 brain template (2-5 months), (d) WM myelinated region identified by EM, (e) KFDA and (f) PVE classifications of the template with masked myelinated WM. The CSF appears white.}
\label{last}
\end{figure}

\begin{figure}[h!]
\centering
\includegraphics[width=2.8cm,height=3.3cm]{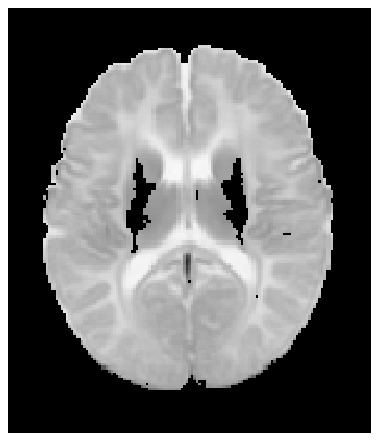}
\includegraphics[width=2.8cm,height=3.3cm]{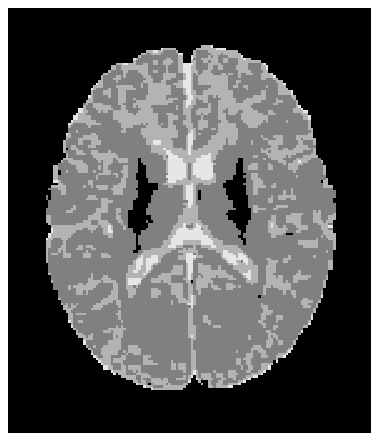}
\includegraphics[width=2.8cm,height=3.3cm]{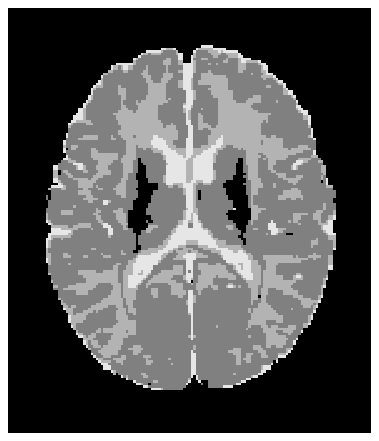}
\\
(a)\hspace{2.5cm}(b)\hspace{2.6cm}(c)\\
\vspace{0.2cm}
\includegraphics[width=2.8cm,height=3.3cm]{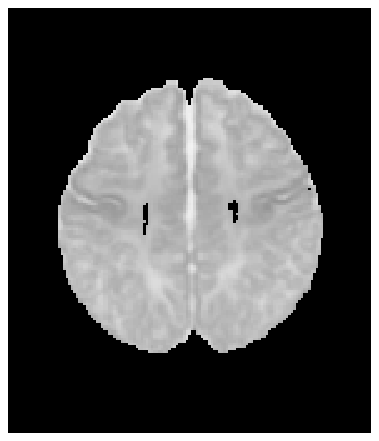}
\includegraphics[width=2.8cm,height=3.3cm]{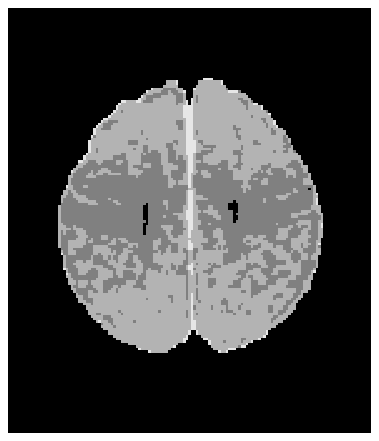}
\includegraphics[width=2.8cm,height=3.3cm]{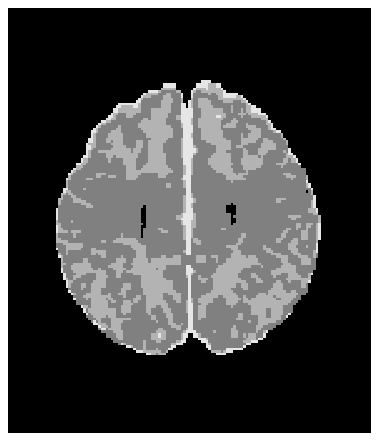}
\\
(d)\hspace{2.5cm}(e)\hspace{2.6cm}(f)\\
\caption{(a),(d) T2w axial slices from an O2 brain template (2-5 months) as unmyelinated WM references, (b),(e) PVE and (c),(f) KFDA classifications of the template with masked myelinated WM. The CSF appears white.}
\label{last2}
\end{figure}

In this paper, we presented a powerful KFDA-based framework that overcomes methodological limitations of existing segmentation approaches in infant brain MRI such as global modelling of tissue intensity distributions and dependency on probabilistic brain atlas. 
\newline
We explored the potential of KFDA in applications to brain tissue classification of infant brain MRI, in particular, the NIH pediatric O2 database. We observed that even with poor initial estimates of the class clusters in the brain template for ages 8 to 11 months KFDA compensates for the underestimates and detects most of the tissues visible in MRI. However, in high contrast subdomains with good initialization (in older children MRI)  KFDA tends to overestimate WM. Overall, application of the KFDA-based method yields a more accurate quantification of brain tissue volumes from infant brain MRI.
 \newline
The proposed method is applicable to brain tissue classification of multichannel brain MRI for ages 8 months and older. We are currently exploring its extension to identification of myelinated and unmyelinated WM in younger infant brains. Figure \ref{last} shows promising preliminary classification results for an O2 template brain for ages 2 to 5 months. The key to successful KFDA-based segmentation of early infancy data lies in the meaningful initialization of brain tissue classes. Initial localization of early myelinated WM regions is a non-trivial task. A label transfer from an older brain of 5 to 8 months of age where WM and GM appear isointense cannot be used to locate regions of early myelination. Also, T1 and T2 intensities of myelinated WM appear to be similar to those of subcortical GM and common classifiers do not capture these intensities correctly.
\newline
However, these data can still be handled in the KFDA framework by subsequent delineation of tissue classes and alteration of reference images for each individual tissue class. For myelinated WM the reference image is the difference between PDw and T1w images that enhances regions of early myelination (Figure \ref{last}.a). For the CSF, the reference is T1w since it provides a higher contrast between the CSF and G+WM. Finally, for unmyelinated WM and GM the reference is T2w (Figures \ref{last}.b-c). 
\newline
Input data also vary depending on what tissue type is being delineated. KFDA is a feature selection method and there is no set dimension to a vector of input data. For myelinated WM segmentation we used only the difference image. For the segmentation of the CSF, all three image types were needed as they all contain information about the CSF. And finally, for the separation of unmyelinated WM and GM, we used T1w and T2w only.
\newline
We first extracted myelinated WM from the reference using an EM algorithm (Figures \ref{last}.d, \ref{last2}.a,\ref{last2}.d). The rest of the brain was initialized using PVE and refined using local KFDA (Figures \ref{last}.e-f, \ref{last2}.c, \ref{last2}.f). 

The local KFDA segmentation approach improves localization of the CSF/GM and GM/WM boundaries in infant brains and facilitates cortical thickness analysis of the O2 dataset.
\newline
The advantage of KFDA extraction of structural information from low contrast images should be exploited in segmentation of lesions associated with Multiple Sclerosis, Mitochondrial Disease or a traumatic brain injury. Given the faint difference between the intensities of WM lesions and GM in T1-weighted images the proposed method can assist in discrimination between WM hyperintensities, WM and GM. It is straightforward to extend KFDA to identification of an extra class of lesions. Using the definition of lesions as outliers in GM and WM clusters \citep{Seghier2008} and voxel categorization into overlapping and outlier sets outliers can then be classified into lesions.
\newline     
It is worth noting that the proposed segmentation by subsequent delineation of tissue classes becomes advantageous for the detection of the lesions located close to the ventricles and the inter-hemispheric fissure. The KFDA discrimination of the CSF before identifying GM and WM will preclude contamination of the lesions by the CSF.
\newline
Most recently, a new high-resolution 3D brain atlas called the BigBrain has been introduced  \citep{Amunts2013} posing new challenges to existing segmentation and validation techniques. It seems natural to use seeded region growing methods \citep{Adams1994} for identification of cortical layers in the BigBrain. However, seed generation is not automated and different orders of processing pixels during region growing process lead to different segmentations. A KFDA-based approach to the BigBrain segmentation is a more powerful alternative that should be explored  in the quest for automated and robust solutions. In this application it can be viewed as a multi-scale approach given its binary tree structure, that is, data segmentation into two classes at each tree node (cortical layer at a lower resolution) in a top-down direction. By moving along the binary tree from coarser to finer scales and carrying down a number of classes the KFDA procedure can be applied to each class until all distinguishable cortical layers are identified. 
\newline
Furthermore, in regards to validation, the traditional approach via Dice coefficients is not yet feasible for the BigBrain since it requires creation of the ground truth by means of a labour-intensive manual segmentation. But in view of our new philosophy of brain segmentation guided by SSIM the BigBrain is the ground truth or reference in itself. Thus, the SSIM-based validation method has potential applications to the BigBrain data segmentation. 
\section{Conclusion}
Experiments presented in this paper show that global tissue classification in young child brain MRI poorly detects WM in the temporal lobe near the base of the brain and cerebellum and significantly underestimates the CSF throughout the brain. The proposed local KFDA-based approach handles highly variable tissue intensities in the O2 data and improves the accuracy of classification in low contrast subdomains.
\newline
This cutting edge atlas-free methodology will advance research in the field of early brain development as well as extend its applications to brain pathology. 

\section{Acknowledgements}
This project was funded in whole or in part by the Montreal Neurological Institute in the form of a postdoctoral fellowship, the National Institute of Child Health and Human Development, the National Institute on Drug Abuse, the National Institute of Mental Health, and the National Institute of Neurological Disorders and Stroke (Contract \#s N01-HD02-3343, N01-MH9-0002, and N01-NS-9-2314, -2315, -2316, -2317, -2319 and -2320). This manuscript expresses the views of the authors and may not reflect the opinions or views of the NIH. 

\bibliographystyle{apalike}
\bibliography{mybiblio}
\nocite{*}
%% not want to use model4-names.bst.


\end{document}